\documentclass[aps,prl,twocolumn,showpacs,superscriptaddress,floatfix]{revtex4}

\usepackage{graphicx}

\begin{document}

\title{Photoluminescence of Tetrahedral Quantum Dot Quantum Wells}

\author{Vladimir A. Fonoberov}
\affiliation{Laboratory of Physics of Multilayer Structures,
Department of Theoretical Physics, State University of Moldova, A.
Mateevici 60, MD-2009 Chi{\c s}in{\u a}u, Moldova}
\affiliation{\parbox{15.5cm}{Nano-Device Laboratory, Department of
Electrical Engineering, University of California--Riverside,
Riverside,} California 92521-0425}

\author{Evghenii P. Pokatilov}
\affiliation{Laboratory of Physics of Multilayer Structures,
Department of Theoretical Physics, State University of Moldova, A.
Mateevici 60, MD-2009 Chi{\c s}in{\u a}u, Moldova}

\author{Vladimir M. Fomin}
\altaffiliation[Also at: ]{TU Eindhoven, P.~O. Box 513,
5600 MB Eindhoven, The Netherlands}
\affiliation{\parbox{17.5cm}{Theoretische Fysica van de Vaste Stoffen,
Departement Natuurkunde, Universiteit Antwerpen,
Universiteitsplein~1,} B-2610 Antwerpen, Belgi\"e}
\affiliation{Laboratory of Physics of Multilayer Structures,
Department of Theoretical Physics, State University of Moldova, A.
Mateevici 60, MD-2009 Chi{\c s}in{\u a}u, Moldova}

\author{Jozef T. Devreese}
\altaffiliation[Also at: ]{TU Eindhoven, P.~O. Box 513,
5600 MB Eindhoven, The Netherlands}
\affiliation{\parbox{17.5cm}{Theoretische Fysica van de Vaste Stoffen,
Departement Natuurkunde, Universiteit Antwerpen,
Universiteitsplein~1,} B-2610 Antwerpen, Belgi\"e}

\date{\today}

\begin{abstract}
Taking into account the tetrahedral shape of a quantum dot quantum well (QDQW) when describing excitonic states, phonon modes and the exciton-phonon interaction in the structure, we obtain within a non-adiabatic approach a quantitative interpretation of the photoluminescence spectrum of a single CdS/HgS/CdS QDQW. We find that the exciton ground state in a tetrahedral QDQW is bright, in contrast to the dark ground state for a spherical QDQW. The position of the phonon peaks in the photoluminescence spectrum is attributed to interface optical phonons. We also show that the experimental value of the Huang-Rhys parameter can be obtained only within the nonadiabatic theory of phonon-assisted transitions.

\end{abstract}


\maketitle

\begin{figure*}[t]
\centering
\begin{tabular}{cc}
\parbox[c]{13.5cm}{\includegraphics[scale=0.68]{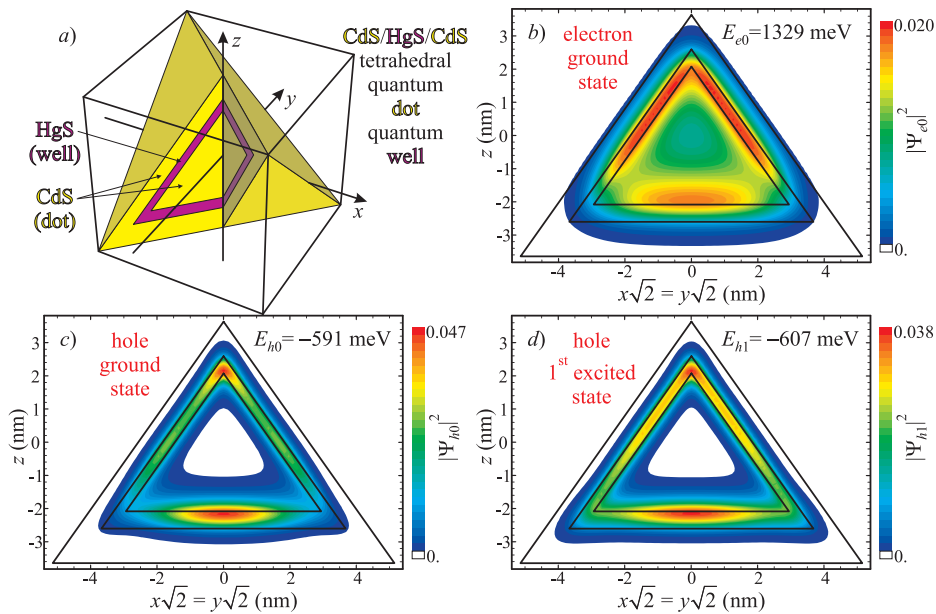}} &
\parbox[c]{4cm}{\caption{(color). The tetrahedral quantum dot quantum well (a) and the average probability densities in the plane $x=y$ of an electron on the
two-fold degenerate ground state level $E_{e0}$ (b) and a hole on
the four-fold degenerate ground state and first excited state levels
$E_{h0}$~(c) and $E_{h1}$ (d). A quarter ($y<x<-y$) of the
tetrahedron in panel (a) is removed to show the position of the
HgS shell. The factor $\sqrt{2}$ multiplying the $x$- and $y$-coordinates in
panels (b)-(d) is used to keep the same scale with the 
$z$-coordinate.}\label{fig:1}}
\end{tabular}
\end{figure*}

Multilayer quantum dots (QD's) are of broad interest not only due to the striking tunability of their physical properties, but also because of the intensively investigated prospects for applications in optoelectronics, microscopy and biology \cite{Alivisatos,Bruchez}. A quantum dot quantum well (QDQW) has been fabricated \cite{Eychmuller} by including a layer of HgS into a nanocrystal of CdS, which has a bandgap larger than that of HgS. The excitons in
a QDQW are mainly confined to the HgS well, which results in a high quantum
yield and an improved photochemical stability. A detailed experimental investigation of the crystalline structure, the shape, and the optical spectra can be carried out on a same single QD \cite{Koberling1}. High-resolution
transmission electron microscopy has revealed that QDQW's are preferentially
tetrahedral particles [see Fig.~\ref{fig:1}(a)] with a zinc blende crystal
lattice \cite{Mews}. Using confocal
optical microscopy, photoluminescence (PL) from a single QDQW has
been measured \cite{Koberling2}. Only the lowest excitonic
transitions in spherical QDQW's have been studied theoretically 
\cite{Jaskolski,Pokatilov1,Devreese,PerezConde,Xie,BJ2003} until now. 
In the present letter, we report the first quantitative interpretation of
the PL spectrum of a QDQW. 

To obtain electron and hole energy spectra in spherical QDQW's,
\textbf{\textit{k}}$\cdot$\textbf{\textit{p}}
\cite{Jaskolski,Pokatilov1,Devreese} and tight-binding
\cite{PerezConde,Xie,BJ2003} calculations have been carried out. Multiband $%
\mathbf{k\cdot p}$ studies for spherical QDQW's yield $s$-type electron and $%
p$-type hole ground states \cite{Jaskolski,Pokatilov1}, leading to the dark
exciton ground state \cite{Devreese}. 
It has been shown that both multiband $\mathbf{k\cdot p}$ and
tight-binding calculations provide a similar qualitative physical
description for the lowest states in QDQW's \cite{BJ2003}.
Recently, we used Burt's
envelope-function representation to develop an 8-band theory which is
particularly suitable for QD heterostructures with thin shells, consisting of
materials with notably different effective-mass parameters and narrow
bandgaps \cite{Pokatilov2}. This multiband theory was successfully applied
to spherical QDQW's \cite{Pokatilov1}, and it is used in the present letter
to study tetrahedral QDQW's.

Experiments show that the PL spectra of single QDQW's do not change
appreciably for different temperatures (1.4 K -- 50 K),
excitation wavelengths (442 nm -- 633 nm) and intensities (0.02
kW/cm$^2$ -- 50 kW/cm$^2$) \cite{Koberling2}. Therefore, the
spectrum corresponds to the equilibrium PL from the exciton ground state $%
\beta _{0}$. In the low-temperature limit, and using the dipole
approximation, the intensities of the phonon lines in the PL spectrum of
QD's are given by \cite{Fomin}: 
\begin{equation}\label{I}
I(\Omega) \propto \sum_{K=0}^\infty
\sum_{\lambda_1,\dots,\lambda_K}
F_{\beta_0,K}^{(\lambda_1,\dots,\lambda_K)}
\delta\Big(\Omega_{\beta_0}-\sum_{j=1}^K
\omega_{\lambda_j}-\Omega\Big),
\end{equation}
where $K$ is the number of a phonon sideband and $\lambda_j$ labels the 
phonon modes with frequencies $\omega_{\lambda_j}$. The general
expression for the amplitudes $F$ of the phonon lines is given in Ref. \cite%
{Fomin}. For the zero-phonon line ($K=0$) and for the one-phonon lines ($K=1$%
), these amplitudes are given by 
\begin{eqnarray}
F_{\beta_0,0}=|d_{\beta_0}|^2,\hspace{5.52cm}\label{F0f}\\
F_{\beta_0,1}^{(\lambda_1)}=\sum_{\beta_1,\beta_2}
\frac{\hbar^{-2} d_{\beta_1}^* d_{\beta_2}
\langle\beta_1|\gamma_{\lambda_1}|\beta_0\rangle
\langle\beta_0|\gamma_{\lambda_1}^*|\beta_2\rangle}{
(\Omega_{\beta_0}-\Omega_{\beta_1}-\omega_{\lambda_1})
(\Omega_{\beta_0}-\Omega_{\beta_2}-\omega_{\lambda_1})},\,\label{F1}
\end{eqnarray}
where $\gamma _{\lambda _{j}}$ are the amplitudes of the exciton-phonon
interaction. The dipole matrix element $d_{\beta }$ is given by 
\begin{equation}\label{d}
d_\beta \propto \int
\delta(\textbf{r}_e-\textbf{r}_h)\Big(\textbf{e},\hat{\textbf{p}} \Big)
\Psi_{exc}^{(\beta)}(\textbf{r}_e,\textbf{r}_h) d\textbf{r}_e
d\textbf{r}_h,
\end{equation}
where $\mathbf{e}$ denotes the light polarization vector, $\hat{\mathbf{p}}$
is the momentum operator \cite{Efros1996}. $\Psi_{exc}^{(\beta)}$ is the wave function for an exciton state $\beta $ with energy $\hbar \Omega _{\beta }$; it is a sum of products of envelope and Bloch functions. For symmetric QD's (e.g. spherical or tetrahedral) the dependence of the dipole matrix element (\ref{d}) on $\mathbf{e}$ reduces to a constant multiplier, which does not affect the relative intensities of the PL peaks.

It is convenient to choose the origin of the coordinate system in the center
of the QDQW [see Fig.~\ref{fig:1}(a)], and to introduce the coordinate $\xi
=(\mathbf{n}_{f},\mathbf{r})$, where $\mathbf{n}_{f}$ is the outward unit vector of the normal to a tetrahedron's facet that contains the point with radius-vector $\mathbf{r}$. The iso-surface of constant $\xi $ is a tetrahedron circumscribed around a sphere with radius $\xi $. In this letter we consider a CdS/HgS/CdS QDQW with the dimensions determined in Refs. \cite{Mews,Koberling2}, namely a CdS core, a HgS shell, and an outer CdS shell bounded by tetrahedrons with $\xi _{1}=1.2$ nm, $\xi _{2}=1.5$ nm, and $\xi _{3}=2.1$ nm. The thicknesses of the HgS shell and of the outer CdS shells are $\xi _{2}-\xi _{1}=0.3$ nm (1 monolayer) and $\xi _{3}-\xi _{2}=0.6$ nm (2 monolayers).

The exciton energies and wave functions have been derived numerically from the
Hamiltonian: 
\begin{equation}
\hat{H}_{exc}=[\hat{H}_{e}+V_{\mathrm{s}\text{-}\mathrm{a}}(\mathbf{r}%
_{e})]-[\hat{H}_{h}-V_{\mathrm{s}\text{-}\mathrm{a}}(\mathbf{r}_{h})]+V_{%
\mathrm{int}}(\mathbf{r}_{e},\mathbf{r}_{h}),  \label{Hexc}
\end{equation}%
where $\hat{H}_{e}$ is the 2-band electron Hamiltonian and $\hat{H}_{h}$ is
the 6-band hole Hamiltonian which we derived from a general 8-band
Hamiltonian for heterostructures \cite{Pokatilov2}. The dielectric mismatch
between QDQW shells leads to the electron and hole
self-interaction potential $V_{\mathrm{s}\text{-}\mathrm{a}}$ and
significantly changes the potential of the electron-hole interaction $V_{%
\mathrm{int}}$ \cite{Fonoberov}. The effective-mass parameters for CdS are taken
from Ref. \cite{Fonoberov}, the dielectric constants from Ref. \cite{Zhang},
and all other required parameters from Ref. \cite{Pokatilov1}. The 
effective-mass parameters for HgS entering the Hamiltonians $\hat{H}_{e}$ and $\hat{H}_{h}$ are
calculated as a function of the energies of the electron and the hole ground
states. The numerical calculation is carried out using a finite-difference
method on the same cubic mesh, both inside and outside the QDQW, as for tetrahedral CdS QD's \cite{Fonoberov}. A mesh length of $0.05%
\sqrt{3}$ nm ensures a relative error for the exciton ground state energy less than 1\%.

HgS is a quantum well for both electrons and holes in the QDQW. While only a
few electron states lie below the CdS bulk conduction band edge and
are localized near the HgS shell, many hole states are trapped in the HgS
shell and lie above the CdS bulk valence band edge. It is seen from Figs.~%
\ref{fig:1}(b) and (c) that the electron in the ground state is practically
uniformly distributed in the HgS shell while the hole in the ground state is
localized near the edges of the tetrahedral HgS shell. Fig.~\ref{fig:1}(d)
shows that, unlike in the ground state, the hole in the first excited state
substantially penetrates the region of facets of the HgS shell, where the
electron density in the ground state is high. 

\begin{figure}[t]
\centering
\includegraphics[scale=0.97]{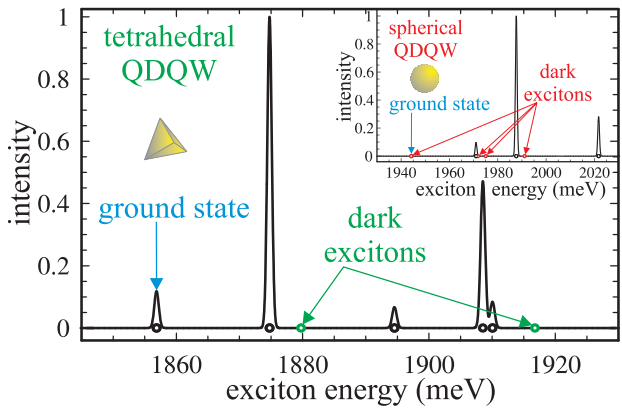}
\caption{(color). Normalized intensities of the lowest exciton
levels in the optical absorption spectrum of a tetrahedral QDQW (the
insert gives the intensities for the spherical QDQW considered in
Ref. \cite{Pokatilov1}). Spectral lines are broadened by a
Gaussian to enhance visualization.} \label{fig:2}
\end{figure}

In the QDQW's optical absorption spectrum, the intensity of a transition to the
exciton level $\beta $ is $|d_{\beta }|^{2}$. Fig.~\ref{fig:2} shows that
the intensity of the second exciton level is $\sim$8.5 times larger than that of
the ground state level. Since the second exciton state is absorbing and the exciton ground state is luminescing, a Stokes shift should be observed between the frequencies of the excitation and the zero-phonon equilibrium PL. In addition to the fact that our calculated exciton ground state
energy of 1857 meV is in excellent agreement with the experimental data of Refs.~\cite{Mews,Koberling2}, the obtained excitation energy of 18 meV is very
close to the value 19 meV of the Stokes shift found in the fluorescence line
narrowing spectrum \cite{Mews}. The insert in Fig.~\ref{fig:2} shows the
result of an analogous finite-difference calculation of the exciton states
for a spherical QDQW \cite{Pokatilov1} with the same thickness of the HgS
shell as that for the tetrahedral QDQW. The exciton ground state is
eight-fold degenerate for both spherical and tetrahedral QDQW's. The
remarkable effect of the QDQW's shape is that the exciton in the ground
state is bright for the tetrahedral QDQW, while it is dark for the spherical QDQW. We find that the ratio of the intensity of the exciton ground state to the
intensity of the first excited state is very sensitive to the shape of
the QDQW's. This sensitivity determines the crucial role of the shape of QDQW's
also for their PL spectra.

\begin{figure}[t]
\centering
\includegraphics[scale=0.92]{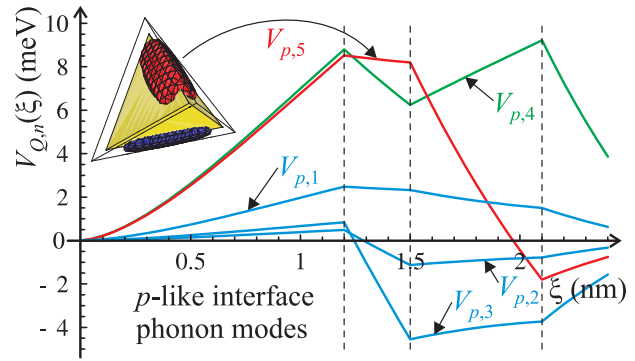}
\caption{(color). Amplitudes of several interface phonon modes
averaged as $V_{Q,n}(\xi)=\Phi_{Q,n}(\xi)
\sqrt{\int_{S(\xi)}\left. \left| \varphi_Q \right|^2 dS
\right/\int_{S(\xi)} dS}$ (for denotations see text). Vertical
dashed lines show the boundaries of the tetrahedral shells. Iso-surfaces $V_{p_z,5}(\mathbf{r})=10$ meV (red) and
$V_{p_z,5}(\mathbf{r})=-10$ meV (blue) in the HgS shell are shown.
The amplitudes $V_{p,4}$ and $V_{p,5}$ give the highest one-phonon
peaks in the PL spectrum in Fig.~\ref{fig:4}.} \label{fig:3}
\end{figure}

The amplitudes of the exciton-phonon interaction are $\gamma _{\lambda }(%
\mathbf{r}_{e},\mathbf{r}_{h})\equiv V_{\lambda }(\mathbf{r}_{e})-V_{\lambda
}(\mathbf{r}_{h})$. In the dielectric continuum approach, the amplitudes $%
V_{\lambda }(\mathbf{r}_{e})$ of the electron-phonon interaction, $V_{\lambda }(\mathbf{r}_{h})$ of the hole-phonon interaction and the phonon
frequency $\omega _{\lambda }$ are solutions of the eigenvalue problem for the
equation: 
\begin{equation}
\nabla \Big(\varepsilon (\omega _{\lambda })\nabla V_{\lambda }\Big)=0\quad 
\text{with}\quad \varepsilon (\omega )=\varepsilon _{\infty }\frac{\omega
^{2}-\omega _{\mathrm{LO}}^{2}}{\omega ^{2}-\omega _{\mathrm{TO}}^{2}},
\label{EqEp}
\end{equation}%
where, for each shell, $\omega _{\mathrm{LO}}$ and $\omega _{\mathrm{TO}%
}\equiv \omega _{\mathrm{LO}}\sqrt{\varepsilon _{\infty }/\varepsilon _{0}}$
are the frequencies of the bulk longitudinal optical (LO) and the transverse
optical (TO) phonons, $\varepsilon _{0}$ and $\varepsilon _{\infty }$ are
the static and the high-frequency dielectric constants, respectively. There
exist two kinds of solutions of Eq.~(\ref{EqEp}). Solutions of the first kind [%
$\varepsilon (\omega _{\lambda })=0$] describe bulk-like phonon modes \cite%
{Klimin}. In the $k$-th shell they have the frequency $\omega _{\mathrm{LO}%
}^{(k)}$ and their amplitude is chosen to satisfy the equation $\quad \nabla
^{2}V_{\lambda }=-q_{k}^{2}V_{\lambda }$ with boundary conditions $\left.
V_{\lambda }\right\vert _{\xi \notin \lbrack \xi _{k-1},\xi _{k}]}=0$. The
number of bulk-like phonon modes is restricted by the condition $q_{k}\leq
\pi /a$, where $a$ is the
lattice constant. Solutions of the second kind [$\varepsilon (\omega
_{\lambda })\neq 0$] describe interface phonon modes. 
To the best of our knowledge, the phonons for the tetrahedral QDQW are theoretically derived here for the first time. We propose for the tetrahedral QDQW an approximation with a spatial behavior close to that for spherical QD's, i.e. $s$-, $p$-like, etc. This means that we seek interface phonon modes as a product of a function $\Phi (\xi )$, which is constant on the tetrahedral surface $S(\xi )$, and a function $\varphi (\mathbf{r})$, which satisfies the equation $\nabla ^{2}\varphi _{Q}=-q^{2}\varphi _{Q}$ and which is zero at infinity. If this method is applied to a spherical QDQW, where $\xi \equiv r$, $\Phi (\xi )$ turns out to be a radial function and $\varphi (\mathbf{r})$ explicitly contains the spherical harmonics that define the spatial symmetry. Substituting 
$V_{\lambda }=\Phi (\xi )\,\varphi_{Q}(\mathbf{r})$ into Eq.~(\ref{EqEp}), multiplying by $\varphi _{Q}^{\ast }$ and integrating over the tetrahedral surface $S(\xi )$, we obtain a
one-dimensional problem with eigenfrequencies $\omega _{Q,n}$ and
eigenfunctions $\Phi _{Q,n}(\xi )$. We used the frequencies of the LO
phonons found in Ref. \cite{Mews}. For the numerical calculation of the bulk-like and interface phonon modes we have used the same finite-difference scheme as described above. For any spatial symmetry of $\varphi _{Q}$ we found five different functions $\Phi _{Q,n}$ that correspond to five eigenfrequencies $\omega _{Q,n}\,(n=1,\dots ,5)$ of the interface phonons. Fig.~\ref{fig:3} shows the amplitudes of the
interface phonon modes $V_{Q,n}$ with $p$-like spatial symmetry, averaged
over the tetrahedral surface $S(\xi )$. In contrast to the amplitudes of the
bulk-like phonon modes which vanish at the boundaries, the amplitudes of the
interface phonon modes have their extremal values at some of the boundaries.

\begin{figure}[t]
\centering
\includegraphics[scale=0.97]{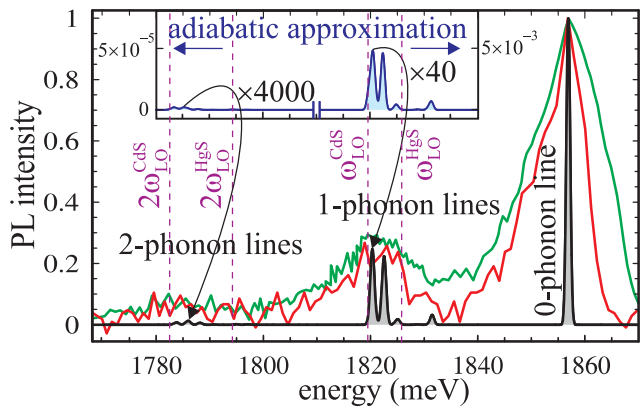}
\caption{(color). PL spectrum of a tetrahedral QDQW normalized to
the zero-phonon line's intensity. Calculated spectral lines (black
curve) are broadened by a Gaussian to facilitate visualization.
Experimental spectra are measured at $T=10$ K with excitation
wavelength 633 nm and intensity 15 kW/cm$^2$ (green curve)
\cite{Koberling2} and with excitation wavelength 442 nm and
intensity 5 kW/cm$^2$ (red curve) \cite{Basche}. The insert gives
one- and two-phonon bands (magnified by factors 40 and 4000,
respectively) as calculated within the adiabatic approximation.}
\label{fig:4}
\end{figure}

In Fig.~\ref{fig:4} the PL spectrum calculated using Eq.~(\ref{I}) for a
single tetrahedral QDQW is compared with two experimental spectra, measured at
different excitation wavelengths and intensities \cite{Koberling2,Basche}.
Most likely spectral diffusion, which has been explained recently by a
reorganization of local electric fields in or around the particles~\cite%
{Neuhauser}, is responsible for the relatively broad lines in the PL spectra of
a single QDQW. The experimentally determined Huang-Rhys parameter is $0.25\pm 0.05$ \cite{Koberling2}. Because of the localization of the exciton in the HgS well, the amplitudes of the bulk-like phonon modes are much smaller than those of the interface phonon modes. The dominant contribution to the intensity of the
one-phonon lines in the calculated PL spectrum is due to $p$-like interface
phonon modes. The two most intense one-phonon peaks correspond to the modes with
amplitudes $V_{p,4}$ and $V_{p,5}$ in Fig.~\ref{fig:3} and energies $34.3$
meV and $36.7$ meV, respectively. The average energy, weighted with the intensities of the two peaks, is $35.5$ meV, in a good agreement with the experimentally determined value of $35.3\pm 0.6$ meV\cite{Koberling2}. 
The intensity of the calculated two-phonon lines cannot directly be compared with that in the experimental spectra because of the large background in this
spectral region. The insert in Fig.~\ref{fig:4} shows the one- and
two-phonon bands in the PL spectrum calculated within the {\it adiabatic
approximation}. In this approximation, the intensities of the one-phonon peaks are defined by Eq.~(\ref{F1}), however, without the summation over the states $\beta_1$ and $\beta_2$, which are replaced by $\beta_0$. Since the exciton ground state $\beta _{0}$ is eight-fold degenerate and its intensity $|d_{\beta _{0}}|^{2}$ is small as compared to the intensities of the higher lying exciton states (see Fig.~\ref{fig:2}), the intensities of the phonon peaks in the adiabatic approximation are dramatically lower than those calculated within the nonadiabatic theory \cite{Fomin}.

In summary, we have shown a strong difference between the optical spectra of spherical and tetrahedral QDQW's. Only the simultaneous
consideration of the tetrahedral shape of a QDQW, interface
optical phonons, and nonadiabatic phonon-assisted transitions
allows for profound understanding of the optical response of a
QDQW. In particular, this analysis allowed us to interpret
the observed PL spectra of a CdS/HgS/CdS QDQW.

\begin{acknowledgments}

The authors thank T. Basch\'e for fruitful
communications and for the provided experimental data. Discussions
with F. Brosens, V. N. Gladilin, S. N. Klimin, and A. A. Balandin are gratefully acknowledged. E. P. P. and V. A. F. acknowledge with
gratitude the kind hospitality received during their visit to the Universiteit Antwerpen (UIA).
This work has been supported by the GOA BOF UA 2000, IUAP, FWO-V
projects G.0274.01, G.0435.03, the W.O.G. WO.025.99N (Belgium),
the EC GROWTH Programme, NANOMAT project G5RD-CT-2001-00545, and
the MRDA-CRDF Award MP2-3044 (Moldova).

\end{acknowledgments}

\end{document}